\documentclass[aps,prl,floatfix,footinbib]{revtex4}
\usepackage{graphicx}
\usepackage{amsmath,amssymb}
\usepackage{subfigure}
\usepackage{bm}
\begin{document}

\title{Perspectives in spintronics: magnetic resonant tunneling, spin-orbit coupling, and GaMnAs}
\author{C. Ertler, A. Matos-Abiague, M. Gmitra, M. Turek, and J. Fabian}
\address{Institute for Theoretical Physics, University of Regensburg, 93040 Regensburg, Germany\\
 }

\begin{abstract}
Spintronics has attracted wide attention by promising novel
functionalities derived from both the electron charge and spin.
While branching into new areas and creating new themes over the past
years, the principal goals remain the spin and magnetic control of
the electrical properties---essentially the I-V
characteristics---and vice versa. There are great challenges ahead
to meet these goals. One challenge is to find niche applications for
ferromagnetic semiconductors, such as GaMnAs. Another is to develop
further the science of hybrid ferromagnetic metal/semiconductor
heterostructures, as alternatives to all-semiconductor {\it room
temperature} spintronics. Here we present our representative recent
efforts to address such challenges. We show how to make a digital
magnetoresistor by combining two magnetic resonant diodes, or how
introducing ferromagnetic semiconductors as active regions in
resonant tunneling diodes leads to novel effects of digital
magnetoresistance and of magnetoelectric current oscillations. We
also discuss the phenomenon of tunneling anisotropic
magnetoresistance in Fe/GaAs junctions by introducing the concept of
the spin-orbit coupling field, as an analog of such fields in
all-semiconductor junctions. Finally, we look at fundamental
electronic and optical properties of GaMnAs by employing reasonable
tight-binding models to study disorder effects.
\end{abstract}



\maketitle

\section{Introduction}

Spintronics is an interdisciplinary field driven by goals of
technological innovations and fundamental discoveries. The two
aspects, applied and fundamental, are intimately intertwined,
pushing spintronics research forward. Thus far, the only important
technological achievements of spintronics remain the giant
magnetoresistance and the tunenling magnetoresistance, both based on
metallic heterostructures \cite{Zutic2004:RMP}. While semiconductor
spintronics has attracted great deal of attention, most of its
device structures are still in the proposal stage, with only few
device principles experimentally demonstrated \cite{Fabian2007:APS}.
Moving ahead with realizing potentially useful spintronic devices is
a major task.

Spintronics has also generated a great deal of fundamental
knowledge. We now have firm understanding of the principles of spin
injection, spin transport, and spin relaxation; we have learned how
to manipulate spins from bulk to nanoscales, or how to bring
ferromagnetism into useful semiconductors \cite{Fabian2007:APS}.
Important, we have learned the limitations of the field, sometimes
fundamental in nature, sometimes technology limited.

This article illustrates spintronics perspectives by discussing
selected topics, from quantum spintronic devices, through spin-orbit
coupling in heterostructures and its implications for tunneling and
magnetocrystalline anisotropy, to physical properties of
ferromagnetic GaMnAs. There are two important questions that these
topics address in their specific forms: (i) What potential
spintronic applications can we think of for ferromagnetic
semiconductors in general, and for GaMnAs in particular?  and (ii)
Are hybrid ferromagnetic metal/semiconductor systems a viable
alternative to all-semiconductor spintronics? The need to answer the
second question stems from the limitations of the physical
properties of useful ferromagnetic semiconductors, such as GaMnAs,
whose ferromagnetism at room temperature remains elusive.

Where do ferromagnetic semiconductors fit? In most proposals for
spintronic devices ferromagnetic semiconductors provide the exchange
splitting to differentiate (mainly transport) properties of spin up
and spin down electrons. Here we discuss our proposal to take the
next step: couple ferromagnetism dynamically with transport and
charging, in resonant diodes. We discuss how such a coupling, in the
quantum well made of a ferromagnetic semiconductor, leads to robust
microwave current oscillations, at nominally dc biases.

One great outcome of spintronics is a good understanding of
spin-orbit coupling in semiconductor interfaces. Due to the current
limitations on the Curie temperature in GaMnAs, we need to consider
conventional sources of spin-polarization: ferromagnetic metals. In
particular, ferromagnetic metal/semiconductor interfaces are
attractive systems for controlling orbital effects by spins. We
discuss here one particular interface, that of Fe/GaAs, and show
that tunneling through such junctions leads to the phenomenon of the
tunneling magnetoresistance (the tunneling resistance depends on the
orientation of the magnetization in the plane perpendicular to the
tunneling), which can be explained by the concept of the spin-orbit
coupling field at such metal/semiconductor interfaces. Both model
and first-principles calculations allow to explore the anisotropies
stemming from this interface induced spin-orbit coupling. We also
point out that while the transport and magnetocrystalline symmetries
both reflect the interface structure, the observed corresponding
anisotropies have different electronic origin.

While finding niche applications for GaMnAs is a nontrivial task,
the material itself is rather fascinating. It is a highly p-doped
semiconductor whose ferromagnetic order is mediated by holes. There
are many fundamental questions still open: the Fermi level lies in
the valence or in the impurity band? What is the effective mass of
the carriers at the Fermi level and what is the mode of the
transport? Are the states at the Fermi level fully extended or
partially localizes? What is the value of the band gap? and so on.
We attempt to answer some of those questions here with two
reasonable tight-binding models, finding that each model explains
certain experimental features in a complementary way, but neither
provides a complete picture; a realistic model of GaMnAs is yet to
be presented.

We first introduce magnetic resonant diodes, then discuss tunneling
anisotropic magnetoresistance and the spin-orbit coupling effects at
ferromagnetic metal/semiconductor interfaces, and finally present
our results and their interpretations of our tight-binding
calculations of large GaMnAs model systems.


\section{Magnetic resonant tunneling devices}

Spintronic devices aim at enhancing the functionality of the
existing electronic technology by exploiting the spin-dependent
properties of solid-state systems. An efficient spin control of the
charge current would allow for the realization of ultimate
magnetoelectric devices, in which information can be processed based
on the spin-state of the system \cite{Zutic2004:RMP}. In this
context, band-engineered magnetic semiconductor heterostructures
offer promising perspectives, since they provide opportunities for
controlling and tuning spin-dependent magneto-transport properties
\cite{Fabian2007:APS}. In particular the rapid development of novel
dilute magnetic semiconductors (DMS) in the last decades
\cite{Ohno1998:S,Dietl:2007,Jungwirth2006:RMP}, in which
semiconductors are made magnetic by doping with transition metal
elements, has considerably enlarged the class of suitable materials
for the growing of heterojunction systems.

Magnetic heterostructures typically consist of stacked layers of
only a few nanometers of both magnetic and nonmagnetic
semiconductors. Along the growth direction the system size becomes
comparable to the phase coherence length of the propagating
carriers, allowing quantum interference effects to govern the
vertical transport properties. According to the band-mismatch of the
different semiconductor materials, some of the layers constitute
energetic barriers for the incident carriers, which could only be
penetrated by tunneling. Nevertheless, in double and multi-barrier
systems transmission probabilities up to nearly 100$\%$ can be found
for certain incident carriers energies, which actually correspond to
the discrete electronic spectra of the formed quantum wells; an
astonishing effect, termed  {\em resonant tunneling}.

From a generic point of view, resonant tunneling structures act like
an energy filter, allowing only carriers with resonant energies to
pass through. The transport becomes strongly sensitive to the
relative alignment of the electronic spectra in the leads and the
wells, which can be modified by external bias voltages. Even small
energy splittings for different spin states in the quantum wells,
e.g., induced by an external magnetic field or an exchange field,
can result in strong ramifications in the transport characteristics,
as has been observed experimentally in paramagnetic
\cite{Slobodskyy2003:PRL} as well as ferromagnetic
\cite{Ohya2007:PRB, Oiwa2004:JMMM} DMSs quantum well structures. Due
to the spin-dependent transmission magnetic resonant tunneling
devices have been proposed for realizing efficient spin valve, spin
filtering, spin switching, and spin detecting devices
\cite{Fabian2007:APS, Petukhov2002:PRL, Ertler2006a:APL}, which form
important building blocks for a spin-based information technology.

In the following we show representative examples of capabilities of
magnetic heterostructures by describing (i) the digital magneto
resistance (DMR), which appears in serially connected resonant
tunneling diodes (RTDs),  and (ii) the dynamic interplay of
transport and magnetic effects in ferromagnetic RTDs, which can lead
to selfsustained current oscillations at a nominally steady bias
voltage.

\begin{figure}
\centering
\subfigure{\includegraphics[width=0.5\linewidth]{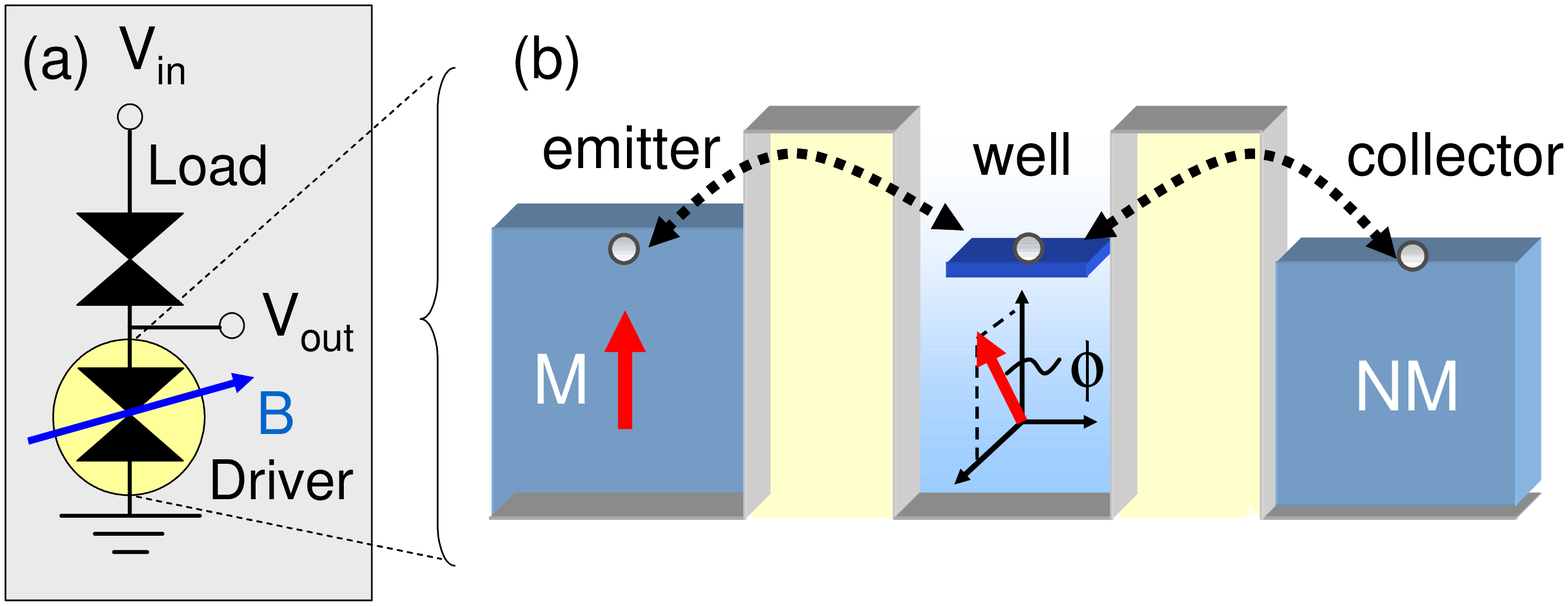}}
\subfigure{\includegraphics[width=0.5\linewidth]{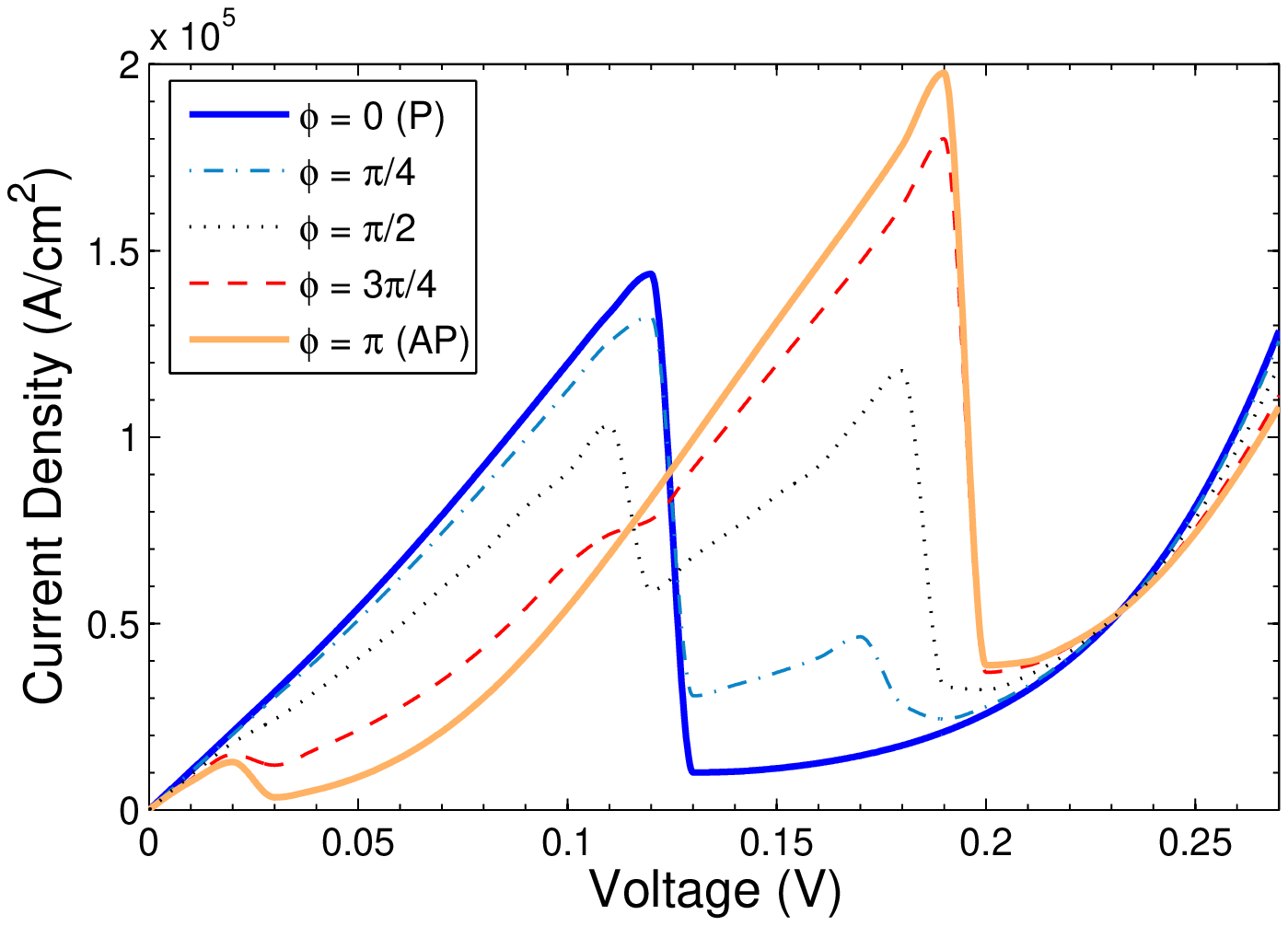}}\\
\caption{\label{fig:mobile}Top panel: (a) circuit diagram of a
magnetic MOBILE. The load is a conventional RTD, whereas the driver
is a RTD with a ferromagnetic emitter and quantum well (b). The well
magnetization can be tilted by an external magnetic field $B$ by the
angle $\phi$. Lower panel: Selfconsistent current-voltage
characteristic of the ferromagnetic driver-RTD for different
orientations of the quantum well magnetization, indicated by the
angle $\phi$.}
\end{figure}

A nice example, which illustrates that the current characteristics
can be strongly modified by changing the magnetic properties of
resonant tunneling structures, is given by a double barrier
structures comprising a ferromagnetic quantum well and a
ferromagnetic emitter, as illustrated in Fig.~\ref{fig:mobile}(b).
Assuming that the magnetization of the quantum well is ''soft`` the
relative alignment of the emitter and well magnetization directions
can be changed by external magnetic fields. Selfconsistent numerical
simulations for GaMnN-like systems \cite{Ertler2007a:PRB} reveal a
strong dependence of the current-voltage (IV) characteristics on the
relative magnetization angle, as shown in the lower panel of
Fig.~\ref{fig:mobile}. In particular, the peak current is
considerably decreased if the the well magnetization is tilted away
from the parallel alignment with the emitter's magnetization. This
effect can be used to realize the proposed DMR-effect
\cite{Ertler2006b:APL, Ertler2007a:PRB} in so-called
monostable-bistable logic elements (MOBILEs) \cite{Maezawa1993:JJAP,
Maezawa:2003}.

A conventional MOBILE consists of two serially connected RTDs, a
load and a driver, as shown in Fig.~\ref{fig:mobile}(a). The
principle of the device operation relies on the nonlinear
IV-characteristics of RTDs. By drawing the circuit load line diagram
a single monostable output voltage for low input voltages is
obtained. However, at high input voltages a bistable working point
regime appears. Which of the two working points is realized actually
depends on the difference of the load and driver peak currents: if
the load peak current is smaller (higher) than the driver's peak
current than the output voltage is low (high). In a nutshell: a
MOBILE can compare the peak currents of the load and driver RTDs,
yielding a digital binary output signal.

In a magnetic variant of the MOBILE the driver is replaced by a
magnetic RTD, e.g, of the kind as discussed above. If the quantum
well magnetization is tilted by an external magnetic field the
magnetic driver's peak current is continuously decreased. At some
critical angle the driver's peak current becomes smaller than the
load peak current, which results in a discrete jump of the output
voltage from low to high. This effectively realizes a {\em digital
magneto resistance}: the total device resistance makes a
discontinuous step by continuously changing the magnetic properties
of the structure, i.e, in this case the tilting angle of the quantum
well magnetization.

The DMR-like effect has, in fact, already been experimentally
demonstrated in a somewhat different device configuration, in which
a metallic giant magneto resistance (GMR) element is shunted to a
nonmagnetic driver diode \cite{Hanbicki2001:APL}. In terms of
applications DMR-devices could be used as very fast read heads of
hard disks, in which the magnetically stored information is directly
converted into a binary electrical signal. Furthermore,
ferromagnetic RTDs can have the advantage to be nonvolatile upon the
loss of power, since the state of the device is stored in the
magnetization direction of particular layers of the structure. Such
nonvolatile devices are attractive for fast random access memory
applications or for reprogrammable logics, in which the logical
circuit function is altered by changing the magnetic state of the
device.

\begin{figure}
\centering
\subfigure{\includegraphics[width=0.5\linewidth]{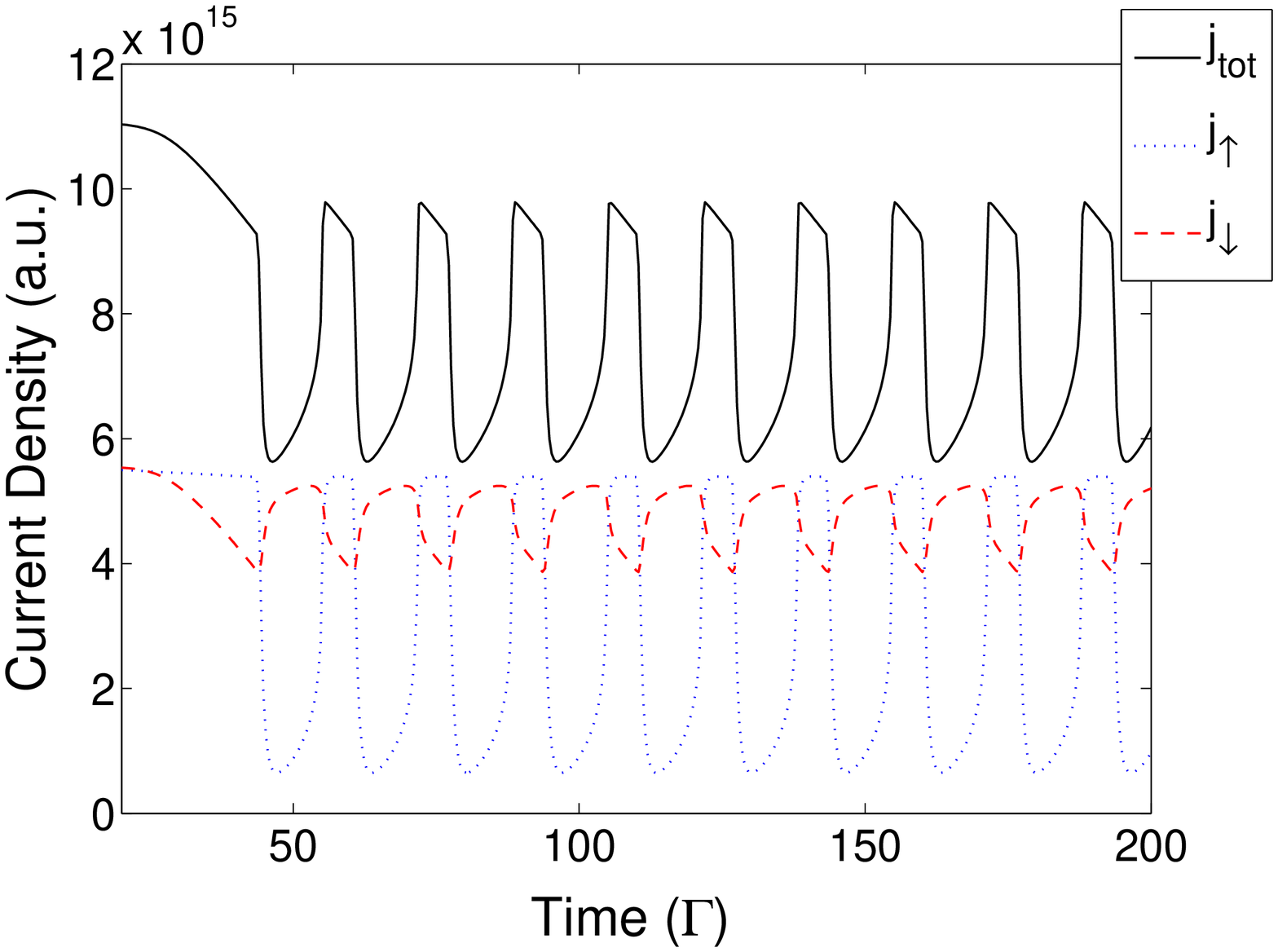}}\\
\subfigure{\includegraphics[width=0.5\linewidth]{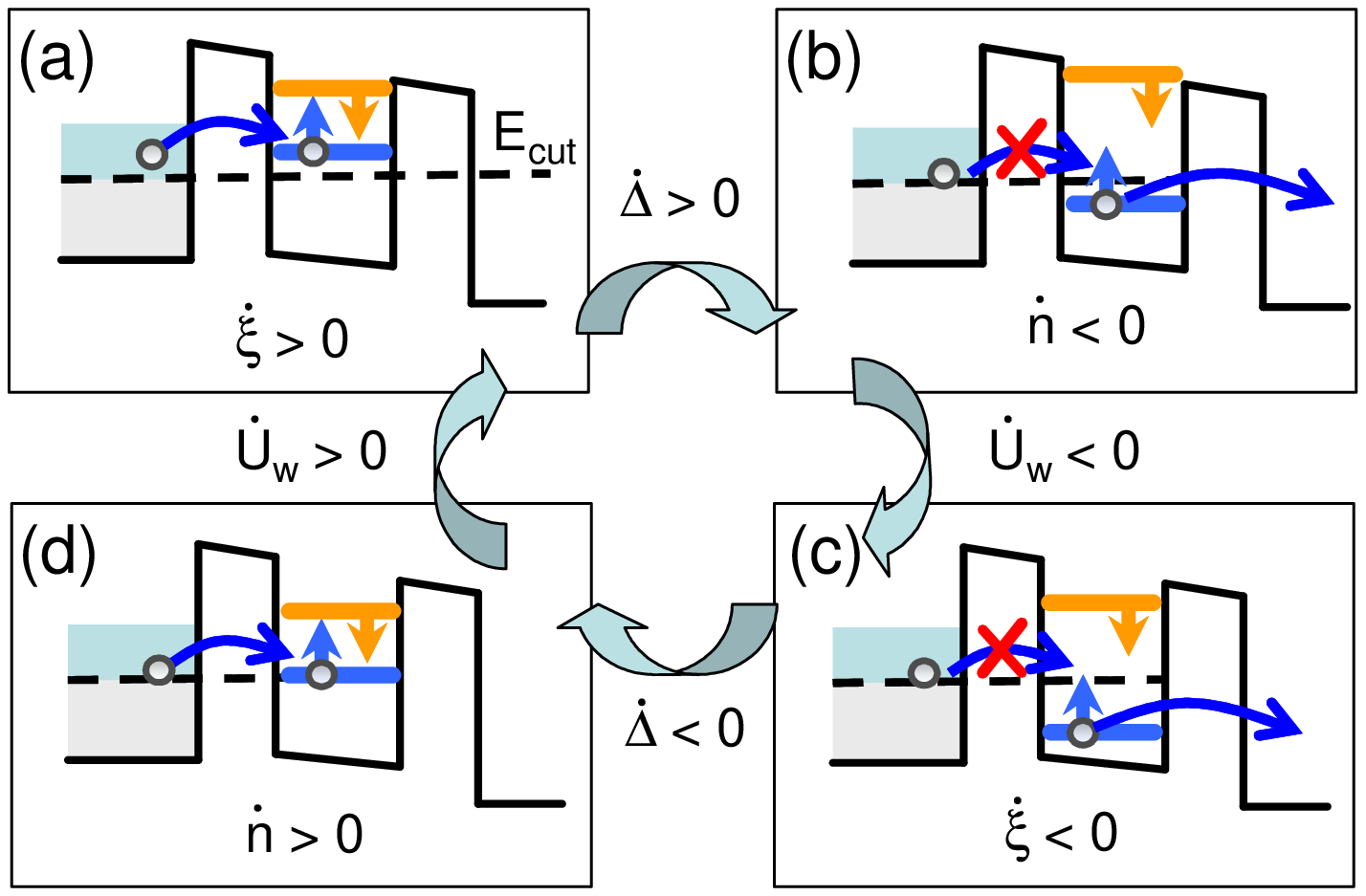}}
\caption{\label{fig:dynQW}Top panel: transients of the oscillating
spin-resolved current density at a fixed dc-bias. Lower panel:
qualitative explanation for the occurrence of self-sustained
oscillations. (a) The in-tunneling spin-up carriers increase the
spin polarization $\xi$, which causes an increasing well splitting
$\Delta$. (b) When the spin up level crosses the cutoff energy the
total particle number $n$ and consequently the electrostatic
potential $U_w$ decreases. (c) This pushes the spin-up level even
deeper into the cut-off region leading to a fast decrease of the
spin polarization and, hence, of the well splitting, bringing the
spin up level back to the emitter's supply region (d), where the
whole process restarts again.}
\end{figure}

In addition to  magnetically tunable steady state properties,
interesting {\em dynamic} phenomena can occur in magnetic
heterostructures according to a peculiar feedback process, in which
the magnetic ordering of the quantum well affects back the resonant
tunneling current.  In DMSs the ferromagnetic order is mediated by
the itinerant carriers, which gives rise to a close interdependence
of the transport and magnetic properties. Indeed, several
experiments succeeded in generating ferromagnetism by tailoring the
carrier density by electrical or optical means \cite{Ohno2000:N,
Boukari2002:PRL}.

In resonant tunneling structures the particle density in the well
strongly depends on the resonant tunneling conditions, which implies
that the critical temperature of magnetic quantum wells should be
tuneable by external biases, as theoretically predicted in the
framework of a mean field model of the well magnetism
\cite{Lee2002:SST, Ganguly2005:PRB}. These studies revealed that the
exchange splitting of the carriers subbands essentially depends (i)
on the spin polarization of the particle density in the well and
(ii) on the overlap of the subband wave function with the magnetic
impurity density profile.

Based on these findings we recently proposed that the combined
nonlinear feedback of Coulomb and magnetic interaction can lead to
self-sustained high-frequency oscillations of the tunneling current
and the quantum well magnetization over a large window of nominally
dc-bias voltages \cite{Ertler2008:condmat}, as illustrated in the
upper panel of Fig.~\ref{fig:dynQW}. The occurrence of these
magnetoelectric oscillations needs a built-in energy cut-off
$E_{\mathrm{cut}}$ of the emitter tunneling rate. This might be
realized, e.g, by a cascaded left barrier, which exponentially
suppresses the tunneling of carriers with energies smaller than
$E_\mathrm{cut}$.

A qualitative discussion and explanation for the appearance of the
self-sustained oscillations is given in the lower panel of
Fig.~\ref{fig:dynQW}: the electrostatic feedback behaves like an
''inertia``, allowing the spin level with energy $E_s$ to get deeper
into the region, in which no carriers can tunnel from the emitter
side anymore ($E_s<E_\mathrm{cut}$), whereas the magnetic feedback
is required to push the spin level back into the emitter's supply
region ($E_s > E_\mathrm{cut}$). The frequency of the oscillations
can be modified by the applied dc-bias voltage, which suggest
applications of ferromagnetic RTDs as tunable high-power
oscillators.

These two examples of magnetic resonant tunneling structures, mainly
chosen on the basis of the authors personal research interest,
illustrate that magnetic heterostructures have a promising potential
for realizing fully integrated magnetoelectronic devices as well as
for gaining a deeper understanding of the mechanism of
carrier-mediated ferromagnetism. Especially the possibilities of
ferromagnetic multi-barrier systems, which naturally provide a rich
variety of compositional configurations, are still largely
unexplored.


\section{Tunneling anisotropic magnetoresistance in magnetic junctions}
\label{TAMR-intro}

\begin{figure}
\includegraphics[width=9cm]{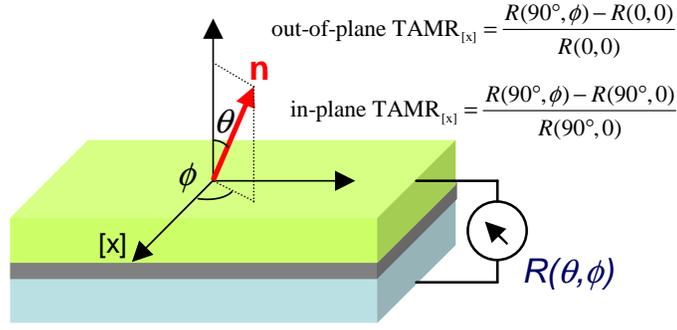}
\caption{Schematics of a TAMR MTJ with a single magnetic layer (the
top layer). The vector $\mathbf{n}$ points in the direction of the
magnetization and [x] denotes the reference axis for $\phi$.}
\label{matos-fig1}
\end{figure}

Spin valve devices are layered heterostructures combining ferro- and
paramagnetic layers. Such devices are at the heart of the
spintronics and are important for the investigation and measurement
of spin injection, spin relaxation, and spin polarization. The spin
valve devices find also multiple applications in the design of read
head sensors and data storage devices, such as magnetic random
access memory. The operation of conventional spin valve devices
relies on the dependence of the magnetoresistance on the relative
orientation of the magnetization in the ferromagnetic leads. The
manifestation of such a phenomenon in tri-layer systems composed of
two magnetic layers separated by a non-magnetic metallic spacer is
called the giant magnetoresistance effect, while in magnetic tunnel
junctions (MTJs) composed of an insulating or semiconducting barrier
sandwiched between two magnetic electrodes is referred to as the
tunneling magnetoresistance (TMR) effect. It has been observed
however, that the magnetoresistance in MTJs may also depend on the
absolute orientation of the magnetization in the ferromagnetic leads
\cite{Gould2004:PRL,Ruster2005:PRL,Saito2005:PRL,Brey2004:APL}. This
phenomenon is called the tunneling anisotropic magnetoresistance
(TAMR) effect \cite{Gould2004:PRL,Brey2004:APL}. The tunneling
magnetoresistance in GaMnAs/GaAlAs/GaMnAs tunnel junctions was
theoretically investigated \cite{Brey2004:APL}. These authors
predicted that, as a result of the strong spin-orbit interaction,
the tunneling magnetoresistance depends on the angle between the
current flow direction and the orientation of the electrode
magnetization. Thus, a difference between the tunneling
magnetoresistances in the in-plane (i.e., magnetization in the plane
of the magnetic layers) and out of plane configurations of up to
$6\%$ was predicted for large values of the electrode spin
polarization \cite{Brey2004:APL}. Here we refer to this phenomenon
as the \emph{out-of-plane} TAMR (see Fig.~\ref{matos-fig1}). Recent
{\it ab initio} calculations in Fe/MgO/Fe magnetic tunnel junctions
(MTJs) predict an out-of-plane TAMR ratio of about 44 \%
\cite{Khan2008:JPCM}. On the other hand, the \emph{in-plane} TAMR
effect refers to the changes in the magnetoresistance when the
in-plane magnetization of the ferromagnetic layer(s) is rotated
around the direction of the current flow (see
Fig.~\ref{matos-fig1}).

It is remarkable that the TAMR is present even in MTJs in which only
one of the electrodes is magnetic and the conventional TMR is
absent. Thus, a new generation of TAMR-based devices could have some
advantages with respect to the devices whose operation is based on
the TMR effect. In contrast to the conventional TMR-based devices,
which require two magnetic layers for their operation, TAMR-based
devices can, in principle, operate with a single magnetic lead. Thus
the TAMR effect could open new possibilities and functionalities for
the operation of spintronic devices. The TAMR effect may have an
influence on the spin-injection from a ferromagnet into a
non-magnetic semiconductor. Therefore, in order to correctly
interpret the results of spin-injection experiments in a spin-valve
configuration, it is essential to understand the nature, properties,
and origin of the TAMR effect.

The first experimental observation of the TAMR effect was performed
in (Ga,Mn)As/AlOx/Au heterojunctions, where an in-plane TAMR ratio
of about $2.7\%$ was found \cite{Gould2004:PRL}. Experimental
investigations of the in-plane TAMR in (Ga,Mn)As/GaAs/(Ga,Mn)As
\cite{Ruster2005:PRL} and (Ga,Mn)As/ZnSe/(Ga,Mn)As
\cite{Saito2005:PRL} tunnel junctions in which both electrodes are
ferromagnetic have also been reported. In the case of
(Ga,Mn)As/ZnSe/(Ga,Mn)As the in-plane TAMR ratio was found to
decrease with increasing temperature, from about $10\%$ at 2 K to
$8.5\%$ at 20 K \cite{Saito2005:PRL}. This temperature dependence of
the in-plane TAMR is more dramatic in the case of
(Ga,Mn)As/GaAs/(Ga,Mn)As, for which a TAMR ratio of order of a few
hundred percent at 4 K was amplified to $150\;000\%$ at 1.7 K
\cite{Ruster2005:PRL}. This huge amplification of the in-plane TAMR
was suggested to originate from the opening of an Efros-Shklovskii
gap \cite{Efros1975:JPC} at the Fermi energy when crossing the
metal-insulator transition \cite{Ruster2005:PRL}.
Measurements of the TAMR in
$\textrm{p}^{+}-$(Ga,Mn)As/$\textrm{n}^{+}$-GaAs Esaki diode devices
has also been reported \cite{Ciorga2007:NJP}. In addition to the
investigations involving vertical tunneling devices
the TAMR has also been studied in
break junctions \cite{Bolotin2006:PRL,Burton2007:PRB},
nanoconstrictions \cite{Giddings2005:PRL,Ciorga2007:NJP} and
nanocontacts \cite{Jacob2008:PRB}.

In addition to the currently low Curie temperature ferromagnetic
semiconductors, the TAMR has recently been experimentally
investigated in Fe/GaAs/Au MTJs \cite{Moser2007:PRL}, Co/AlO$_x$/Au
MTJs \cite{Liu2008:NL}, and in multilayer-(Co/Pt)/AlO$_x$/Pt
structures \cite{Park2008:PRL}.

In what follows we shall focus our discussion on the case of the
in-plane TAMR (for brevity we will refer to it just as the TAMR
effect) recently measured in Fe/GaAs/Au MTJs \cite{Moser2007:PRL}.
In such an experiment the magnetization of the ferromagnetic Fe
electrode was rotated in-plane by an angle $\phi$ with respect to
the [110] crystallographic direction and the magnetoresistance
$R(\phi)=R(90^{\circ},\phi)$ measured as a function of $\phi$ (see
Fig.~\ref{matos-fig1}). The strength of the TAMR is given by
\cite{Fabian2007:APS}
\begin{equation}\label{tamr}
    \textrm{TAMR}_{[110]}(\phi)=\frac{R(\phi)-R(0)}{R(0)}.
\end{equation}

\begin{figure}
\includegraphics[width=9cm]{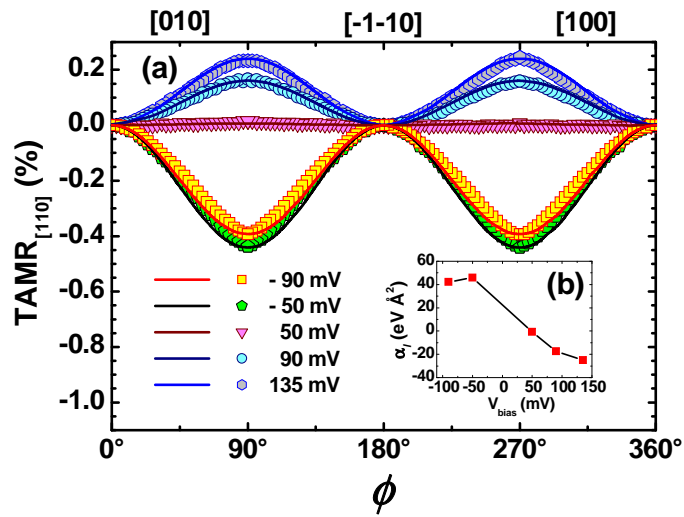}
\caption{(a) Angular dependence of the TAMR in a Fe/GaAs/Au MTJ for
different values of the bias voltage $V_{_\textrm{bias}}$. Solid
lines correspond to theoretical results while symbols represent the
experimental data, as deduced from Ref.~\cite{Moser2007:PRL}. (b)
Extracted bias dependence of $\alpha_{l}$.} \label{matos-fig2}
\end{figure}

The angular dependence of the TAMR in a Fe/GaAs/Au tunnel
heterojunction is shown in Fig.~\ref{matos-fig2}(a) for different
values of the bias voltage $V_{_\textrm{bias}}$. The symbols
represent the experimental data (conveniently mirrored) as deduced
from Ref.~\cite{Moser2007:PRL}. Two fundamental features in the
behavior of the TAMR can be appreciated in Fig.~\ref{matos-fig2}(a):
(i) the experimentally measured TAMR exhibits a two-fold symmetry
with respect to the magnetization direction in the ferromagnet, (ii)
by varying the bias voltage the TAMR can be inverted (i.e., its sign
can be changed). In order to explain these experimental findings we
have proposed a model (for details see Ref.~\cite{Fabian2007:APS})
in which the two-fold symmetry of the TAMR in
ferromagnet/semiconductor/normal metal junctions emerges from the
interference of Dresselhaus and Bychkov-Rashba SOCs.
The combined action of these two spin-orbit interactions leads to
net, anisotropic SOC with a two-fold symmetry which is compatible
with the $C_{2v}$ symmetry of the semiconductor interfaces and
manifest in the tunneling magnetoresistance. The Dresselhaus SOC
results from the bulk inversion asymmetry of the zinc-blende
semiconductor (GaAs in this case) and is characterized by the
Dresselhaus parameter $\gamma$ which has a finite value in the
semiconductor region (for GaAs $\gamma\approx 24$ eV \AA$^2$
\cite{Fabian2007:APS}) and vanishes elsewhere.

For the small voltages considered in the experiment the main
contribution to the Bychkov-Rashba SOC comes from the strong
potential gradients at the interfaces rather than from an external
electric field. The interface Bychkov-Rashba SOC arises due to the
semiconductor interfaces inversion asymmetry \cite{Fabian2007:APS}
and is characterized by the parameters $\alpha_l$ and $\alpha_r$
which measure the strengths of the SOC at the left
(ferromagnet/semiconductors) and right (semiconductor/normal metal)
interfaces, respectively. These parameters are not known for
metal/semiconductor interfaces and have to be extracted from
available experimental data \cite{Moser2007:PRL} or from
first-principles calculations (see the following section). For the
case of an Fe/GaAs/Au MTJ it turns out that the size of the TAMR is
dominated by the parameter $\alpha_{l}$ corresponding to the
ferromagnet/semiconductor interface \cite{Fabian2007:APS}. Since the
values of $\alpha_{r}$ are irrelevant for the TAMR, one can set this
parameter, without loss of generality, to zero. This leaves
$\alpha_{l}$ as a single fitting parameter when comparing to
experiment. The theoretical results are shown by solid lines in
Fig.~\ref{matos-fig2}. The agreement between theory and experiment
is very satisfactory. The values of the phenomenological parameter
$\alpha_{l}$ are determined by fitting the theory to the
experimental value of the TAMR at $\phi=90^{\circ}$. This is enough
for the theoretical model to reproduce the {\emph complete} angular
dependence of the TAMR, demonstrating the robustness of the model.
By performing the fitting procedure for different values of the bias
voltage $V_{_\textrm{bias}}$ the bias dependence of $\alpha_{l}$ can
be extracted. The results are shown in Fig.~\ref{matos-fig2}(b). The
obtained values of $\alpha_{l}$ have reasonable magnitudes, not too
different from known values in semiconductor interfaces
\cite{Fabian2007:APS}. For a bias voltage of about 50 mV the
parameter $\alpha_{l}\approx 0$ and the TAMR vanishes [compare
Figs.~\ref{matos-fig2}(a) and (b)]. At a bias slightly bellow 50 mV
the interface Bychkov-Rashba parameter changes sign, resulting in
the inversion (change of sign) of the TAMR, as seen from
Fig.~\ref{matos-fig2}(a).

In order to understand the main qualitative behavior of the TAMR, a
simplified phenomenological model was proposed in
Refs.~\cite{Moser2007:PRL,Fabian2007:APS}. Such a model is based on
a perturbative expansion of the transmissivity in powers of
$\mathbf{n}.\mathbf{w}$, with $\mathbf{n}$ denoting the unit vector
defining the magnetization direction and
$\mathbf{w}=(-\bar{\alpha_{l}}k_{y}+\bar{\gamma}k_{x},\bar{\alpha_{l}}k_{x}-\bar{\gamma}k_{y},0)$
being the effective spin-orbit coupling field.
Here $\bar{\alpha}$ and $\bar{\gamma}$ represent the average values
of the Bychkov-Rashba and linearized Dresselhaus parameters,
respectively (note that $\bar{\alpha}_{l}\propto \alpha_{l}$ and
$\bar{\gamma}\propto \gamma$). It was then shown, on the basis of
very general symmetry considerations, that up to second order in the
strength of the SOC field the angular dependence of the TAMR is
determined by the relation (for details see
Ref.~\cite{Fabian2007:APS})
\begin{equation}\label{tamr-pheno}
    \textrm{TAMR}_{[110]}\propto \alpha_{l}\gamma [\cos(2\phi)-1],
\end{equation}
which is consistent with the experimental data and the full
theoretical calculations [see Fig.~\ref{matos-fig2}(a)]. One can see
from Eq.~(\ref{tamr-pheno}) that bias-induced changes of the sign of
$\alpha_{l}$ lead to the inversion of the TAMR. It is also clear
that when $\alpha_{l}\gamma=0$, the TAMR is suppressed. Therefore,
for a finite in-plane TAMR both Bychkov-Rashba and Dresselhaus SOCs
must necessarily be present. This is in contrast to the case of the
TAMR in the out-of-plane configuration for which the presence of
only Bychkov-Rashba coupling suffices
\cite{Brey2004:APL,Chantis2007:PRL,Khan2008:JPCM}.
%

\section{\label{MGmitra}Ab-Initio study of spin-orbit coupling effects in Fe/GaAs interfaces}

The spin-orbit interaction couples electron spin to the orbital
moment. In magnetic systems the spin-orbit interaction is the
primary source of the magnetocrystalline anisotropy. In Fe films
grown on semiconductors with a cubic zinc-blende structure, e.g.,
ZnSe, GaAs or InAs, a uniaxial magnetic anisotropy has been reported
\cite{Krebs1987:JAP}. This comes rather as a surprise since both the
Fe and the semiconductors have cubic symmetry. The easy axis has
been found along $[110]$ or $[1\bar{1}0]$ directions. The origin of
the different easy axis orientations has been attributed to the
different semiconductor surface reconstructions. Kneedler {\it et
al.} \cite{Kneedler1997:PRB} showed that strain in the Fe film grown
on GaAs(001) surface has negligible effect on the easy axis
orientation indicating that it is independent of the surface
reconstruction. Moosb\"uhler {\it et al.} \cite{Moosbuehler2002:JAP}
investigated $(4\times 2)$ and $(2\times 6)$ surface reconstructions
and pointed out that easy axis is always along $[110]$ with
practically the same amplitude for different Fe film thicknesses. It
is of importance to introduce a realistic model for the origin of
the uniaxial magnetic anisotropy as well as to relate it to the TAMR
effect, whose bias induced reversal is well described, on the one
hand, by the interference between Bychkov-Rashba and Dresselhaus
spin-orbit couplings \cite{Fabian2007:APS,Moser2007:PRL} and, on the
other hand, by the reversal of the spin polarization
\cite{Chantis2007:PRL99}. The interface structure also affects the
Schottky barrier height \cite{Demchenko2006:PRB} and consequently
spin injection to the GaAs.

\begin{figure}
\includegraphics[width=0.5\columnwidth]{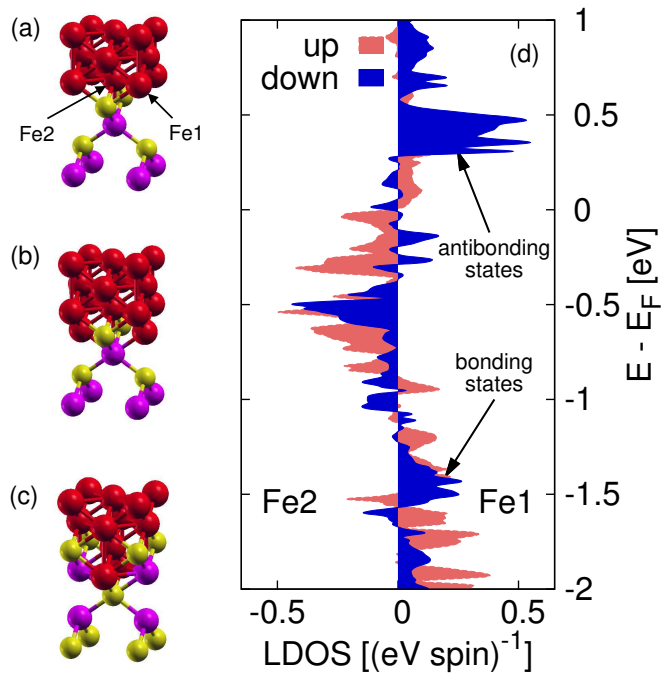}
\caption{Schematics of $1\times 1$ Fe/GaAs interface models
(a)~model~A, (b)~model~B and (c)~model~C. (d)~difference of
$d$-projected local density of states of two nonequivalent Fe atoms
at the flat interface (model~A) for magnetization along
$[1\bar{1}0]$. } \label{Fig:martin:1}
\end{figure}

As the lattice constant of the bulk GaAs ($5.65\,{\rm\AA}$) is
almost twice the lattice constant of the bcc Fe ($2.87\,{\rm\AA}$),
the smooth epitaxial growth of Fe on a GaAs(001) surface is possible
\cite{Krebs1987:JAP,Brockmann1999:JMMM}. Erwin {\it et al.}
\cite{Erwin2002:PRB} studied, using density functional theory, the
stability of ideal $1\times 1$ Fe/GaAs interfaces. Their study
revealed that the thickness of the Fe film as well as the type of
the GaAs surface termination (As- or Ga-like) affects the interface
stability. Three types of $1\times 1$ Fe/GaAs interface models have
been investigated \cite{Erwin2002:PRB}: (i)~model~A - flat interface
[Fig.\ref{Fig:martin:1}(a)]; (ii)~model~B - partially intermixed
interface [Fig.\ref{Fig:martin:1}(b)]; (iii)~model~C - fully
intermixed interface [Fig.\ref{Fig:martin:1}(c)]. When more than two
atomic layers of Fe are deposited on a GaAs(001) surface, the flat
(model A) and partially intermixed (model B) interface models are
more stable than the fully intermixed ones (model C). As the $pd$
hybridization lowers the cohesive energy \cite{Mirbt2003:PRB}, the
antibonding bonds between Fe and As are more stable than the Fe-Ga
bonds. Interestingly, while the density functional calculations
imply that the As-terminated flat (model A) interface is more stable
then the partially intermixed (model B) \cite{Erwin2002:PRB} one, a
recent Z-contrast scanning transmission electron microscopy
experiment reported a single plane of alternating Fe and As atoms
(model B) in an Fe/AlGaAs interface \cite{Zega2006:PRL}.

In the following we discuss electronic properties of an ideal slab
structure of the flat interface model (model A) between Fe and
As-terminated GaAs(001). The slab encased in vacuum of the
$6{\rm\AA}$ thickness contains 9 atomic planes of GaAs and three
atomic planes of bcc Fe on top of the GaAs(001) surface. The
orthorombic symmetry of the system is reduced due to the spin-orbit
coupling, depending on the orientation of the magnetization in the
Fe. The presence of the much debated uniaxial magnetic anisotropy
can be explained in this flat interface model by means of
directional covalent bonds at the interface \cite{Sjostedt2002:PRL}.
The interface breaks the natural continuation of the $sp^3$
tetrahedral bonds of the zinc-blende structure. Because of that, in
what follows two inequivalent groups of Fe atoms at the interface
can be distinguished. The first group contains Fe atoms Fe1 [see
Fig.\ref{Fig:martin:1}(a)], which are in "bonding" positions, e.g.,
the sites where Ga atoms would have been situated. The second group
contains Fe2 atoms that correspond to voids in a zinc-blende
structure. These nonequivalent Fe atoms show similar local densities
of states (LDOS), but if one plots the difference of the
$d$-projected LDOS, see Fig.\ref{Fig:martin:1}(d), the coupling of
the Fe film to the GaAs substrate appears to be clear. The
$d$-states about 0.5~eV below the Fermi level ($E_{\rm F}$) are
split in the case of Fe1 due to bonds between hybridized $sp^3$
states (As) and $d$-states (Fe1) to bonding (1.4~eV bellow $E_{\rm
F}$) and antibonding states (0.4~eV above $E_{\rm F}$). Thus, by
means of LDOS one finds the answer that uniaxial magnetocrystalline
anisotropy has its origin in anisotropic interfacial bonds
\cite{Sjostedt2002:PRL}, which represents a microscopic mechanism
causing the magnetic anisotropy in the Fe/GaAs heterostructure. It
has been shown in case of Zn-terminated Fe/ZnSe(001) system that all
Fe atoms in "bonding" positions contribute to uniaxial magnetic
anisotropy with easy axis along $[1\bar{1}0]$ direction while atoms
in void-like positions along $[110]$ direction with 40-times lower
magnitude \cite{Sjostedt2002:PRL}. In the case of Se-terminated
interface the easy axis points along $[110]$. In the case of Fe/GaAs
heterostructure one assumes similar trends since the splitting of
the bonding states is analogous.

In addition to the magnetic anisotropy, a novel transport phenomenon
has been discovered in Fe/GaAs heterostructures: a tunneling
anisotropic magnetoresistance (TAMR) effect, as discussed in the
previous section. This effect, which has been observed in Fe/GaAs/Au
junctions \cite{Moser2007:PRL} with epitaxial Fe/GaAs interfaces,
has challenged our understanding of ferromagnetic
metal/semiconductor junctions. Indeed, a typical spin-valve like
behavior has been observed for a junction that contains only a
single ferromagnetic layer; at least two layers are needed for
conventional spin valve phenomena. To investigate the origin of the
TAMR effect have proposed a simplified model of a
ferromagnet/semiconductor heterostructure based on general symmetry
considerations \cite{Moser2007:PRL, Fabian2007:APS}. The spin-orbit
coupling in the junction is manifested through linear (in momentum)
Bychkov-Rashba and Dresselhaus terms, which reflect, respectively,
the structure and bulk inversion asymmetry of the Fe/GaAs
heterostructure. Interference of these spin-orbit couplings leads in
[001] quantum wells to the $C_{2v}$ symmetry. The same symmetry then
appears in the tunneling magnetoresistance, which is a function of
the magnetization direction \cite{Moser2007:PRL}. In addition, we
consider a spin splitting, which describes the splitting of the
corresponding spin states due to the strong exchange splitting in
the Fe film. The eigenspectrum, in the limit of a  strong exchange
splitting  (well justified since the splitting in Fe is about
1.5~eV, much more than the splitting due to the spin-orbit
interaction) close to the Brillouin zone center is
\begin{equation}
\epsilon_n^\sigma(k) = \epsilon_n^{(0)} - \sigma \big(
k\left[\alpha_n \sin(\phi-\theta) + \gamma_n
\sin(\phi+\theta)\right]\big) \,. \label{Eq:martin}
\end{equation}
Since we have allowed for band dependent Bychkov-Rashba ($\alpha_n$)
and Dresselhaus ($\gamma_n$) parameters, we can identify the effect
of the spin-orbit coupling on the parabolic band $\epsilon_n^{(0)}$
as a shift that depends on the spin ($\sigma=\pm 1$), the sign of
$\alpha_n$ and $\gamma_n$, the orientation of the magnetization
($\phi$) as well as the orientation of the momentum ($\theta$).
Performing a fitting of ab-initio data to the eigenspectrum
Eq.(\ref{Eq:martin}) we have obtained effective Bychkov-Rashba and
Dresselhaus parameters. We have found that the Bychkov-Rashba
parameters change sign within the energy window of about 250~meV
above and below $E_{\rm F}$. Since the sign of TAMR depends on the
product $\alpha\gamma$ [see Eq.(\ref{tamr-pheno})] we can conclude
that the bias induced TAMR reversal, observed in Ref.
\cite{Moser2007:PRL}, can be explained by the change of the sign of
the Bychkov-Rashba parameters.

It would be reasonable to expect that the magnetocrystalline
anisotropy can be influenced by an applied bias in the same way as
the tunneling anisotropic magnetoresistance. As we have seen,
though, the two effects originate from different electronic states:
magnetocrystalline anisotropy is given by the states deep below the
Fermi level, while TAMR is determined by the propagating states
close to the Fermi level. The bias dependence of the TAMR means that
states within a transport window (given by the bias) contribute. As
these states have in principle different spin-orbit field
parameters, upon averaging over the transport window we obtain
different resistance at different biases; we are probing different
states. Such an averaging would not change the magnetocrystalline
anisotropy which is dominated by the spin-orbit effects of the deep
lying states. However, one could expect effects due to the bias
dependent electronic deformation of these states.


\section{Tight-binding modeling of electronic and optical properties of GaMnAs}

\begin{figure}[b]
\resizebox{0.3\textwidth}{!}{\includegraphics{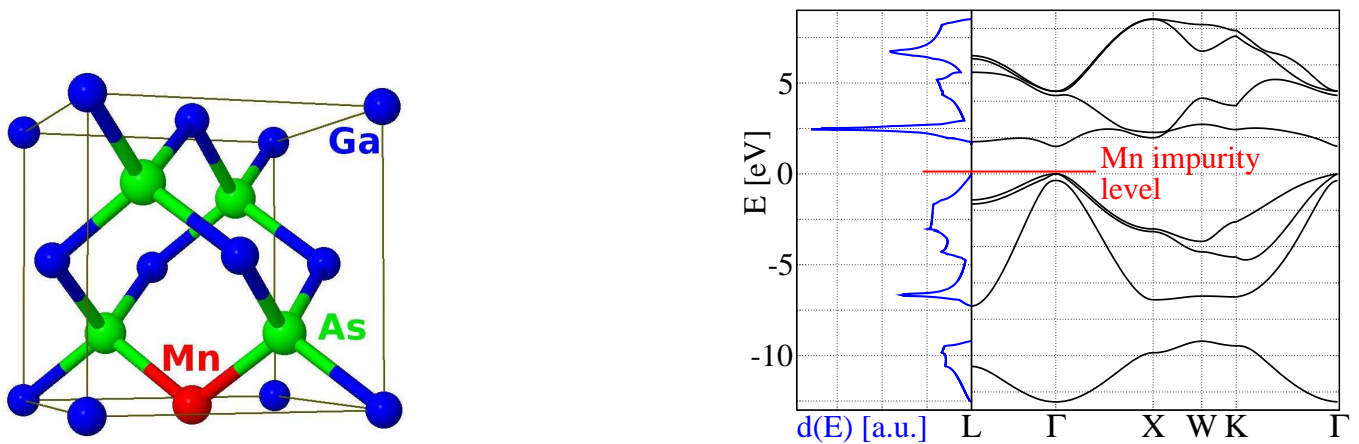}}
\hfill
\resizebox{0.45\textwidth}{!}{\includegraphics{fig_GaAs_BsDos_TalwarTing.eps}}
\caption{\label{fig:GaMnAs}Crystal structure of GaAs with
substitutional Mn impurity (left panel) and a calculated
tight-binding band structure and the density of states (right
panel).}
\end{figure}

Diluted magnetic semiconductors have been subject to very active
research during the past years due to their promising role for new
spin-electronic devices \cite{Fabian2007:APS}. GaMnAs is a much
studied material system among these magnetic semiconductors as its
Curie temperature is rather high, reaching about 170 K
\cite{Ohno1998:S, Jungwirth2006:RMP}. The host material, GaAs, is a
III-V fcc semiconductor with a band gap of $1.5$~eV; the Mn impurity
has a binding energy of $0.11$~eV when it replaces a Ga atom, see
Fig.~\ref{fig:GaMnAs}. Since Mn is short one $p$ electron compared
with Ga at the $4sp$ valence levels, substitutional Mn impurities
act as acceptors that carry a local magnetic moment leading to a
hole mediated ferromagnetism \cite{MacDonald05}.

Despite efforts to understand the bulk properties of ferromagnetic
GaMnAs both from an experimental and theoretical point of view a
number of questions remain open. It is still an unsettled issue
whether the free carriers (holes) reside in an impurity band which
is detached from the host valence band or in the valence band
itself. To this end various absorption experiments \cite {Nagai01,
Hirakawa02, Burch06a} were performed. However, the Mn concentration
threshold for which the impurity band, that appears for very low Mn
concentrations, eventually merges with the host valence band could
not uniquely be determined. Measurements of the Fermi energy in
GaMnAs and the size of the gap as a function of the impurity
concentration are presented in \cite{Tsuruoka02, Thomas07}. These
characteristics are important for the design and study of spintronic
devices as well as for spin injection experiments. Further open
questions concern the scattering and localization properties of the
holes caused by the disorder in the material. Besides the impurity
band picture \cite{Burch06a}, an interpretation based on a merged
impurity and valence band together with the existence of localized
states in the band tail could also explain the absorption
experiments \cite{Jungwirth07a}. A wide range of theoretical models
describing the electronic properties of GaMnAs has been reviewed in
\cite{Jungwirth2006:RMP}. Among these are first-principles
calculations \cite{Sandratskii04,Stroppa08}, effective single
particle tight-binding approaches
\cite{Tang04a,Dorpe05,Sankowski06,Masek07a}, tight-binding
approaches in combination with percolation theory
\cite{Bhatt02,Kaminski02,DasSarma03}, dynamical mean field theories
\cite{Majidi06}, effective theories based on $k \cdot p$
Hamiltonians \cite{Dietl00,Dietl01,Yang03,Elsen07}, and large-scale
Monte-Carlo studies of a real space Hamiltonian \cite{Yildrim07a}.

In our studies we address these open questions using a
phenomenological multi-band tight-binding simulation method for
ferromagnetic bulk GaMnAs. This material specific method treats the
Mn impurities in a non-perturbative way and can be applied to system
sizes of up to 2000 atoms (nowadays) with reasonable computational
effort. It is based on well-known $sp^3$ tight-binding models for
the host GaAs material which include spin orbit coupling
\cite{Chadi77a, Talwar82a}. The disorder effects are modeled by
changing the on-site and hopping terms of those Ga sites that are
replaced by the Mn impurities. We compare two recent proposals, one
due to Ma{\v s}ek and the other due to Tang and Flatt\'e. The first
approach is based on a second-nearest neighbor approximation for the
host GaAs \cite{Talwar82a} and determines the Mn tight-binding
parameters by means of first principles calculations in a Mn
concentration regime around $10$\% \cite{Masek07a, Masek07b}. The
second approach remains within the nearest neighbor approximation
\cite{Chadi77a}. It is based on a fit of two physically relevant
tight-binding parameters to reproduce the binding energy of a single
Mn impurity in the GaAs host material \cite{Tang04a, Kitchen06a}. In
either case an effective single-particle model is obtained which
treats the carrier-carrier interactions in a mean field
approximation.

As it is not clear yet which of the two different parameter sets (if
any) is better suited for the description of available experimental
data we have investigated the density of states, the position of the
Fermi energy, the size of the gap, the localization properties of
the wavefunctions and the optical conductivity of GaMnAs for both
models and compared the results to experimental findings
\cite{Turek08a}.

\begin{figure}[t]
\resizebox{0.6\textwidth}{!}{\includegraphics{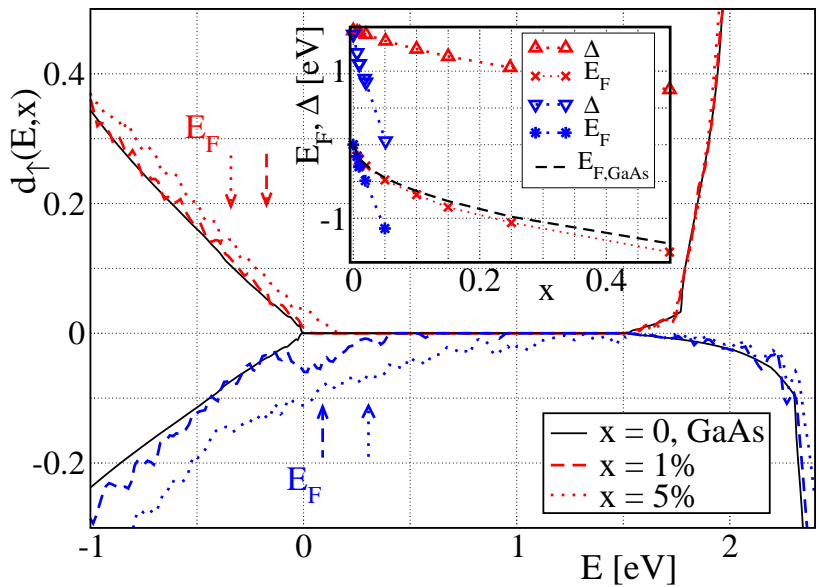}}
\caption{\label{fig:dos_gap_masek_flatte}Density of states for Ma{\v
s}ek's model (positive sign) and Tang, Flatt\'e's model (negative
sign). The solid lines show the corresponding results for clean
GaAs. INSET: Gap size $\Delta$ (triangles: Ma{\v s}ek -- upwards;
Tang, Flatt\'e -- downwards) and Fermi energy $E_{\rm F}$ (Ma{\v
s}ek -- crosses; Tang, Flatt\'e -- stars). The dashed line
represents the Fermi energy of p-GaAs with the corresponding number
of holes added.
}
\end{figure}

First, we present the results of both models for the density of
states in the dependence on the Mn concentration $x$. In
Fig.~\ref{fig:dos_gap_masek_flatte} the density $d(E,x)$ is shown
for the spin up states only as the spin down contributions follow
the GaAs density very closely. Because the two models are based on
different parameter sets for clean GaAs the resulting densities also
show a slightly different behavior, in particular for the conduction
band.

Photoconductivity experiments\cite{Thomas07} on GaMnAs
heterostructures yield a gap $\Delta \approx 1.41$~eV for $x\approx
3$~\% with $E_{\rm F}$ being close to the impurity level. Scanning
tunneling microscopy measurements \cite{Tsuruoka02} resulted in
$\Delta \approx 1.23$~eV for a similar concentration. However, a
quantitative comparison with the experimental data and our results
is not possible due to the uncertainty in the number of Mn
interstitials in the experimental data.

For Ma{\v s}ek's model the density of states shows only minor
modifications due to the increasing disorder, see
Fig.~\ref{fig:dos_gap_masek_flatte}. There is no detached impurity
band forming in the relevant range of Mn concentrations $x$. The
Fermi energy decreases into the host valence band as the Mn
impurities increase the number of holes. If the Fermi energy is
measured from the top of the GaMnAs valence band as a function of
the Mn impurity one finds a close similarity to clean p-GaAs with
the corresponding number of holes added, see inset of
Fig.~\ref{fig:dos_gap_masek_flatte}. For the second model due to
Tang and Flatt\'e the situation is completely different. An impurity
band, which merges with the host valence band at Mn concentrations
$\gtrsim 1$\% can clearly be identified, see lower curves of
Fig.~\ref{fig:dos_gap_masek_flatte}. A large number of valence
states are shifted into the host gap resulting in a Fermi energy
which increases compared to the top of the host valence band. The
gap eventually disappears at $x\approx 5$~\%; this is not consistent
with experiment.


Besides the density of states, the localization properties of the
states around the Fermi energy play an important role for transport
phenomena. We studied these properties by means of the participation
ratio and found that the states around the Fermi energy are extended
in both models. Similar analysis for energies closer to the band
edge leads to a different result for the model by Tang and Flatt\'e.
At these higher energies in the GaMnAs valence band tail one finds
strongly localized states \cite{Turek08a}.

The optical conductivity $\sigma( \omega )$ is closely related to
the absorption and can be directly accessed in
experiments\cite{Burch06a, Katsumoto01, Hirakawa02}. However, there
are still some discrepancies in the results and the interpretation
of the data concerning the low frequency limit, the position of the
intervalence band peaks, and the deduced effective masses and mean
free paths. The two different models we studied result in two
qualitatively different line shapes for $\sigma(\omega )$. Ma{\v
s}ek's model shows a Drude peak at zero frequency while no such peak
is observed in Tang and Flatt\'e's model. However, despite all the
qualitative differences of the two models, i.e. concerning the
existence of an impurity band, the calculated effective masses are
similar being within the range of $0.4$ -- $1.0$ of the free
electron mass.

We conclude that phenomenological tight-binding methods are well
suited for the simulation of material properties of ferromagnetic
semiconductors as they include the disorder in a non-perturbative
way. However, neither of the two proposed parameterizations studied
by us describes satisfactorily available experimental data. They
both describe some properties correctly while failing at others.

\section*{References}
\bibliography{spin}

\providecommand{\newblock}{}
\begin{thebibliography}{10}
\expandafter\ifx\csname url\endcsname\relax
  \def\url#1{{\tt #1}}\fi
\expandafter\ifx\csname urlprefix\endcsname\relax\def\urlprefix{URL }\fi
\providecommand{\eprint}[2][]{\url{#2}}

\bibitem{Zutic2004:RMP}
{\v{Z}uti\'c} I, Fabian J and {Das Sarma} S 2004 {\em Rev. Mod. Phys.\/} {\bf
  76} 323--410

\bibitem{Fabian2007:APS}
Fabian J, Matos-Abiague A, Ertler C, Stano P and {\v{Z}uti\'c} 2007 {\em Acta
  Phys. Slov.\/} {\bf 57} 565--907

\bibitem{Ohno1998:S}
Ohno H 1998 {\em {\sl Science}\/} {\bf 281} 951--956

\bibitem{Dietl:2007}
Dietl T 2007 in W~P{\"o}tz, J~Fabian and U~Hohenester, eds, {\em Modern Aspects
  of Spin Physics\/} (Springer, Berlin) pp 1--46

\bibitem{Jungwirth2006:RMP}
Jungwirth T, Sinova J, Ma\v{s}ek J, Ku\v{c}era J and MacDonald A~H 2006 {\em
  Rev. Mod. Phys.\/} {\bf 78} 809--864

\bibitem{Slobodskyy2003:PRL}
Slobodskyy A, Gould C, Slobodskyy T, Becker C~R, Schmidt G and Molenkamp L~W
  2003 {\em Phys. Rev. Lett.\/} {\bf 90} 246601

\bibitem{Ohya2007:PRB}
Ohya S, Hai P~N, Mizuno Y and Tanaka M 2007 {\em Phys. Rev. B\/} {\bf 75}
  155328--1--155328--6

\bibitem{Oiwa2004:JMMM}
Oiwa A, Moriya R, Kashimura Y and Munekata H 2004 {\em J. Magn. Magn. Mater.\/}
  {\bf 272-276} 2016--2017

\bibitem{Petukhov2002:PRL}
Petukhov A~G, Chantis A~N and Demchenko D~O 2002 {\em Phys. Rev. Lett.\/} {\bf
  89} 107205

\bibitem{Ertler2006a:APL}
Ertler C and Fabian J 2006 {\em Appl. Phys. Lett.\/} {\bf 89}
  242101--1--242101--3

\bibitem{Ertler2007a:PRB}
Ertler C and Fabian J 2007 {\em Phys. Rev. B\/} {\bf 75} 195323--1--195323--9

\bibitem{Ertler2006b:APL}
Ertler C and Fabian J 2006 {\em Appl. Phys. Lett.\/} {\bf 89}
  193507--1--193507--3

\bibitem{Maezawa1993:JJAP}
Maezawa K and Mizutani T 1993 {\em Jpn. J. Appl. Phys.\/} {\bf 32} L42--L44

\bibitem{Maezawa:2003}
Maezawa K and F{\"o}rster A 2003 in R~Waser, ed, {\em Nanoelectronics and
  Information Technology\/} (Wiley-VCH, Weinheim) pp 407--424

\bibitem{Hanbicki2001:APL}
Hanbicki A~T, Mango R, Cheng S~F, Park Y~D, Bracker A~S and Jonker B~T 2001
  {\em Appl. Phys. Lett.\/} {\bf 79} 1190--1192

\bibitem{Ohno2000:N}
Ohno H, Chiba D, Matsukura F, Abe T~O~E, Dietl T, Ohno Y and Ohtani K 2000 {\em
  {\sl Nature}\/} {\bf 408} 944--946

\bibitem{Boukari2002:PRL}
Boukari H, Kossacki P, Bertolini M, Ferrand D, Cibert J, Tatarenko S, Wasiela
  A, Gaj J~A and Dietl T 2002 {\em Phys. Rev. Lett.\/} {\bf 88} 207204

\bibitem{Lee2002:SST}
Lee B, Jungwirth T and MacDonald A~H 2002 {\em Semi. Sci. Techn.\/} {\bf 17}
  393--403

\bibitem{Ganguly2005:PRB}
Ganguly S, Register L~F, Banerjee S and MacDonald A~H 2005 {\em Phys. Rev. B\/}
  {\bf 71} 245306--1--245306--8

\bibitem{Ertler2008:condmat}
Ertler C and Fabian J 2008 {\em Phys. Rev. Lett.\/} {\bf 101} 077202

\bibitem{Gould2004:PRL}
Gould C, {R\"{u}ster} C, Jungwirth T, Girgis E, Schott G~M, Giraud R, Brunner
  K, Schmidt G and Molenkamp L~W 2004 {\em Phys. Rev. Lett.\/} {\bf 93} 117203

\bibitem{Ruster2005:PRL}
{R\"{u}ster} C, Gould C, Jungwirth T, Sinova J, Schott G~M, Giraud R, Brunner
  K, Schmidt G and Molenkamp L~W 2005 {\em Phys. Rev. Lett.\/} {\bf 94} 027203

\bibitem{Saito2005:PRL}
Saito H, Yuasa S and Ando K 2005 {\em Phys. Rev. Lett.\/} {\bf 95} 086604

\bibitem{Brey2004:APL}
Brey L, Tejedor C and {Fern\'{a}ndez-Rossier} J 2004 {\em Appl. Phys. Lett.\/}
  {\bf 85} 1996--1998

\bibitem{Khan2008:JPCM}
Khan M~N, Henk J and Bruno P 2008 {\em J. Phys. Condens. Matter\/} {\bf 20}
  155208

\bibitem{Efros1975:JPC}
Efros A~L and Shklovskii B~I 1975 {\em J. Phys. C\/} {\bf 8} L49--L51

\bibitem{Ciorga2007:NJP}
Ciorga M, Schlapps M, Einwanger A, {Gei\ss ler} S, Sadowski J, Wegscheider W
  and Weiss D 2007 {\em New J. Phys.\/} {\bf 9} 351

\bibitem{Bolotin2006:PRL}
Bolotin K~I, Kuemmeth F and Ralph D~C 2006 {\em Phys. Rev. Lett.\/} {\bf 97}
  127202

\bibitem{Burton2007:PRB}
Burton J~D, Sabrinov R~F, Velev J~P, Mryasov O~N and Tsymball E~Y 2007 {\em
  Phys. Rev. B\/} {\bf 76} 144430

\bibitem{Giddings2005:PRL}
Giddings A~D, Khalid M~N, Jungwirth T, Wunderlich J, Yasin S, Edmonds
  R~P~C~K~W, Sinova J, Ito K, {K-Y Wang}, Williams D, Gallagher B~L and Foxon
  C~T 2005 {\em Phys. Rev. Lett.\/} {\bf 94} 127202

\bibitem{Jacob2008:PRB}
Jacob D, {Fern\'{a}ndez-Rossier} J and Palacios J~J 2008 {\em Phys. Rev. B\/}
  {\bf 77} 165412

\bibitem{Moser2007:PRL}
Moser J, Matos-Abiague A, Schuh D, Wegscheider W, Fabian J and Weiss D 2007
  {\em Phys. Rev. Lett.\/} {\bf 99} 056601

\bibitem{Liu2008:NL}
Liu R~S, Michalak L, Canali C~M, Samuelson L and Pettersson H 2008 {\em Nano
  Lett.\/} {\bf 8} 848--852

\bibitem{Park2008:PRL}
Park B~G, Wunderlich J, Williams D~A, Joo S~J, Jung K~Y, Shin K~H,
  {Olejn\'{\i}k} K, Schik A~B and Jungwirth T 2008 {\em Phys. Rev. Lett.\/}
  {\bf 100} 087204

\bibitem{Chantis2007:PRL}
Chantis A~N, Belashchenko K~D, Tsymbal E~Y and {van Schilfgaarde} M 2007 {\em
  Phys. Rev. Lett.\/} {\bf 98} 046601

\bibitem{Krebs1987:JAP}
Krebs J~J, Jonker B~T and Prinz G~A 1987 {\em J. Appl. Phys.\/} {\bf 61} 3744

\bibitem{Kneedler1997:PRB}
Kneedler E~M, Jonker B~T, Thibado P~M, Wagner R~J, Shanabrook B~V and Whitman
  L~J 1997 {\em Phys. Rev. B\/} {\bf 56} 8163 -- 8168

\bibitem{Moosbuehler2002:JAP}
Moosb{\" u}hler R, Bensch F, Dumm M and Bayreuther G 2002 {\em J. Appl.
  Phys.\/} {\bf 91} 8757

\bibitem{Chantis2007:PRL99}
Chantis A~N, Belashchenko K~D, Smith D~L, Tsymbal E~Y, {van Schilfgaarde} M and
  Albers R~C 2007 {\em Phys. Rev. Lett.\/} {\bf 99} 196603

\bibitem{Demchenko2006:PRB}
Demchenko D~O and Liu A~Y 2006 {\em Phys. Rev. B\/} {\bf 73} 115332

\bibitem{Brockmann1999:JMMM}
Brockmann M, Z{\"o}lfl M, Miethaner M and Bayreuther G 1999 {\em J. Magn. Magn.
  Mater.\/} {\bf 198} 384

\bibitem{Erwin2002:PRB}
Erwin S~C, Lee S and Scheffler M 2002 {\em Phys. Rev. B\/} {\bf 65} 205422

\bibitem{Mirbt2003:PRB}
Mirbt S, Sanyal B, Isheden C and Johansson B 2003 {\em Phys. Rev. B\/} {\bf 67}
  155421

\bibitem{Zega2006:PRL}
Zega T~J, Hanbicki A~T, Erwin S~C, \v{Z}uti\'c I, Kioseoglou G, Li C~H, Jonker
  B~T and Stroud R~M 2006 {\em Phys. Rev. Lett.\/} {\bf 96} 196101

\bibitem{Sjostedt2002:PRL}
Sj\"ostedt E, L~Nordstr\"om F~G and Eriksson O 2002 {\em Phys. Rev. Lett.\/}
  {\bf 89} 267203

\bibitem{MacDonald05}
MacDonald A~H, Schiffer P and Samarth N 2005 {\em Nat. Mat.\/} {\bf 4} 195

\bibitem{Nagai01}
Nagai Y, Kunimoto T, Nagasaka K, Nojiri H, Motokawa M, Matsukura F, Dietl T and
  Ohno H 2001 {\em Jpn. J. Appl. Phys.\/} {\bf 40} 6231

\bibitem{Hirakawa02}
Hirakawa K, Katsumoto S, Hayashi T, Hashimoto Y and Iye Y 2002 {\em Phys. Rev.
  B\/} {\bf 65} 193312

\bibitem{Burch06a}
Burch K~S, Shrekenhamer D~B, Singley E~J, Stephens J, Sheu B~L, Kawakami R~K,
  Schiffer P, Samarth N, Awschalom D~D and Basov D~N 2006 {\em Phys. Rev.
  Lett.\/} {\bf 97} 087208

\bibitem{Tsuruoka02}
Tsuruoka T, Tachikawa N, Ushioda S, Matsukura F, Takamura K and Ohno H 2002
  {\em Applied Physics Letters\/} {\bf 81}(15) 2800--2802

\bibitem{Thomas07}
Thomas O, Makarovsky O, Patan\`{e} A, Eaves L, Campion R~P, Edmonds K~W, Foxon
  C~T and Gallagher B~L 2007 {\em Applied Physics Letters\/} {\bf 90}(8) 082106

\bibitem{Jungwirth07a}
Jungwirth T, Sinova J, MacDonald A~H, Gallagher B~L, Novak V, Edmonds K~W,
  Rushforth A~W, Campion R~P, Foxon C~T, Eaves L, Olejn'ik K, Masek J, Yang
  S~R~E, Wunderlich J, Gould C, Molenkamp L~W, Dietl T and Ohno H 2007 {\em
  Phys. Rev. B\/} {\bf 76} 125206

\bibitem{Sandratskii04}
Sandratskii L~M, Bruno P and Kudrnovsk{\'{y}} J 2004 {\em Phys. Rev. B\/} {\bf
  69} 195203

\bibitem{Stroppa08}
Stroppa A, Picozzi S, Continenza A, Kim M~Y and Freeman A~J 2008 {\em Phys.
  Rev. B\/} {\bf 77}(3) 035208

\bibitem{Tang04a}
Tang J~M and Flatt{\'{e}} M~E 2004 {\em Phys. Rev. Lett.\/} {\bf 92} 047201

\bibitem{Dorpe05}
VanDorpe P, VanRoy W, DeBoeck J, Borghs G, Sankowski P, Kacman P, Majewski J~A
  and Dietl T 2005 {\em Phys. Rev. B\/} {\bf 72}(20) 205322

\bibitem{Sankowski06}
Sankowski P, Kacman P, Majewski J and Dietl T 2006 {\em Physica E\/} {\bf 32}
  375

\bibitem{Masek07a}
Ma{\v{s}}ek J 2007 {\em private communication\/}

\bibitem{Bhatt02}
Bhatt R, Berciu M, Kennett M~P and Wan X 2002 {\em Journal of
  Superconductivity: INM\/} {\bf 15} 71

\bibitem{Kaminski02}
Kaminski A and DasSarma S 2002 {\em Phys. Rev. Lett.\/} {\bf 88} 247202

\bibitem{DasSarma03}
DasSarma S, Hwang E~H and Kaminski A 2003 {\em Phys. Rev. B\/} {\bf 67} 155201

\bibitem{Majidi06}
Majidi M~A, Moreno J, Jarrell M, Fishman R~S and Aryanpour K 2006 {\em Phys.
  Rev. B\/} {\bf 74}(11) 115205

\bibitem{Dietl00}
Dietl T, Ohno H, Matsukura F, Cibert J and Ferrand D 2000 {\em Science\/} {\bf
  287} 1019

\bibitem{Dietl01}
Dietl T, Ohno H and Matsukura F 2001 {\em Phys. Rev. B\/} {\bf 63} 195205

\bibitem{Yang03}
Yang S~R~E, Sinova J, Jungwirth T, Shim Y~P and MacDonald A~H 2003 {\em Phys.
  Rev. B\/} {\bf 67} 045205

\bibitem{Elsen07}
Elsen M, Jaffres H, Mattana R, Tran M, George J~M, Miard A and Lemaitre A 2007
  {\em Phys. Rev. Lett.\/} {\bf 99}(12) 127203

\bibitem{Yildrim07a}
Yildirim Y, Alvarez G, Moreo A and Dagotto E 2007 {\em Phys. Rev. Lett.\/} {\bf
  99} 057207

\bibitem{Chadi77a}
Chadi D~J 1977 {\em Phys. Rev. B\/} {\bf 16} 790

\bibitem{Talwar82a}
Talwar D~N and Ting C~S 1982 {\em Phys. Rev. B\/} {\bf 25} 2660

\bibitem{Masek07b}
Ma{\v{s}}ek J, Kudrnovsk{\'{y}} J, M{\'{a}}ca F, Sinova J, MacDonald A~H,
  Campion R~P, Gallagher B~L and Jungwirth T 2007 {\em Phys. Rev. B\/} {\bf 75}
  045202

\bibitem{Kitchen06a}
Kitchen D, Richardella A, Tang J~M, Flatt{\'{e}} M and Yazdani A 2006 {\em
  Nature\/} {\bf 442} 436

\bibitem{Turek08a}
Turek M, Siewert J and Fabian J 2008 {\em Phys. Rev. B\/} {\bf 78} 085211

\bibitem{Katsumoto01}
Katsumoto S, Hayashi T, Hashimoto Y, Iye Y, Ishiwata Y, Watanabe M, Eguchi R,
  Takeuchi T, Harada Y, Shin S and Hirakawa K 2001 {\em Mater. Sci. Eng. B\/}
  {\bf 84} 88

\end{thebibliography}

\end{document}